# Meme and Variations: A Computer Model of Cultural Evolution

**Liane Gabora**

Holland's (1975) genetic algorithm is a minimal computer model of natural selection that made it possible to investigate the effect of manipulating specific parameters on the evolutionary process. If culture is, like biology, a form of evolution, it should be possible to similarly abstract the underlying skeleton of the process and develop a minimal model of it. Meme and Variations, or MAV, is a computational model, inspired by the genetic algorithm, of how ideas evolve in a society of interacting individuals (Gabora 1995). The name is a pun on the classical music form 'theme and variations', because it is based on the premise that novel ideas are variations of old ones; they result from tweaking or combining existing ideas in new ways (Holland *et al*. 1981). MAV explores the impact of several phenomena that are unique to culture. These are introduced briefly here, and the technical details of how they are implemented will be presented shortly.

The first is knowledge-based operators. Brains detect regularity and build schemas with which they adapt the mental equivalents of mutation and recombination to the constraints of their world, and the situation at hand. Thus they generate novelty strategically, on the basis of past experience. Knowledge-based operators are a crude attempt to incorporate this into the model.

The second crudely implemented cultural phenomenon is imitation. Ideas for how to perform actions spread when agents copy what one another is doing. This enables them to share solutions to the problems they face.

The third is mental simulation; agents can 'imagine', or guess, how successful an idea would be if it were implemented before they actually commit to implementing it. This provides them with a rudimentary form of selection before the phenotypic expression of an idea. Once again, the way this phenomenon is implemented is not terribly life-like, but the goal here was to abstract the essence of the phenomenon and see how it affects the evolution of culture.

Every iteration, each neural-network based agent in an artificial society has the opportunity to acquire a new idea, either through 1) innovation, by changing a previously learned idea, or 2) imitation, by copying what a neighbor is doing. Thus, variation occurs through mutation of pre-existing ideas, selection occurs through choice of which pre-existing idea to mutate, and how to mutate it, and ideas spread through imitation.

## 1. THE MODEL

Since the model is artificially limited with respect to the number and kinds of features an idea can have, it does not hurt in this case to adopt the terminology of biology. Thus the features or components of an idea are referred to as loci, and alternative forms of a locus are referred to as alleles. The processes that generate variation are referred to as mutation operators. Forward mutation is mutation away from the original (or, as biologists refer to it, the wild type) allele, and backmutation is mutation from an alternative form back to the original. An individual is referred to as an agent, and the set of all agents is referred to as the society.

### 1.1 The Domain

Donald (1991) has provided substantial evidence that the earliest culture took the form of physical actions, such as displays of aggression or submission. The ideas in MAV can be thought of as mating displays. An idea is a pattern consisting of six loci that dictate the degree of movement for six body parts: left arm, right arm, left leg, right leg, head, and tail. Each locus has a floating point activation between -0.5 and 0.5 which determines the amount of movement (angle of rotation from rest position) of the corresponding body part when the idea is implemented. A value of 0.0 corresponds to rest position; values above 0.0 correspond to upward movement, and values below 0.0 correspond to downward movement. Floating point loci activations produce graded limb movement. However, for the purpose of mutation, loci are treated as if there are only three possible alleles at each locus: stationary, up, and down. Six loci with three possible alleles each gives a total of 729 possible ideas.

### 1.2 The Neural Network

The neural network is an autoassociator; it learns the identity function between input and output patterns. It has six input/output units numbered 1 through 6, corresponding to the six body parts. It has six hidden units numbered 7 through 12, corresponding to the general concepts, "arms", "legs", "left", "right", "movement", and "symmetry" (Figure 1).

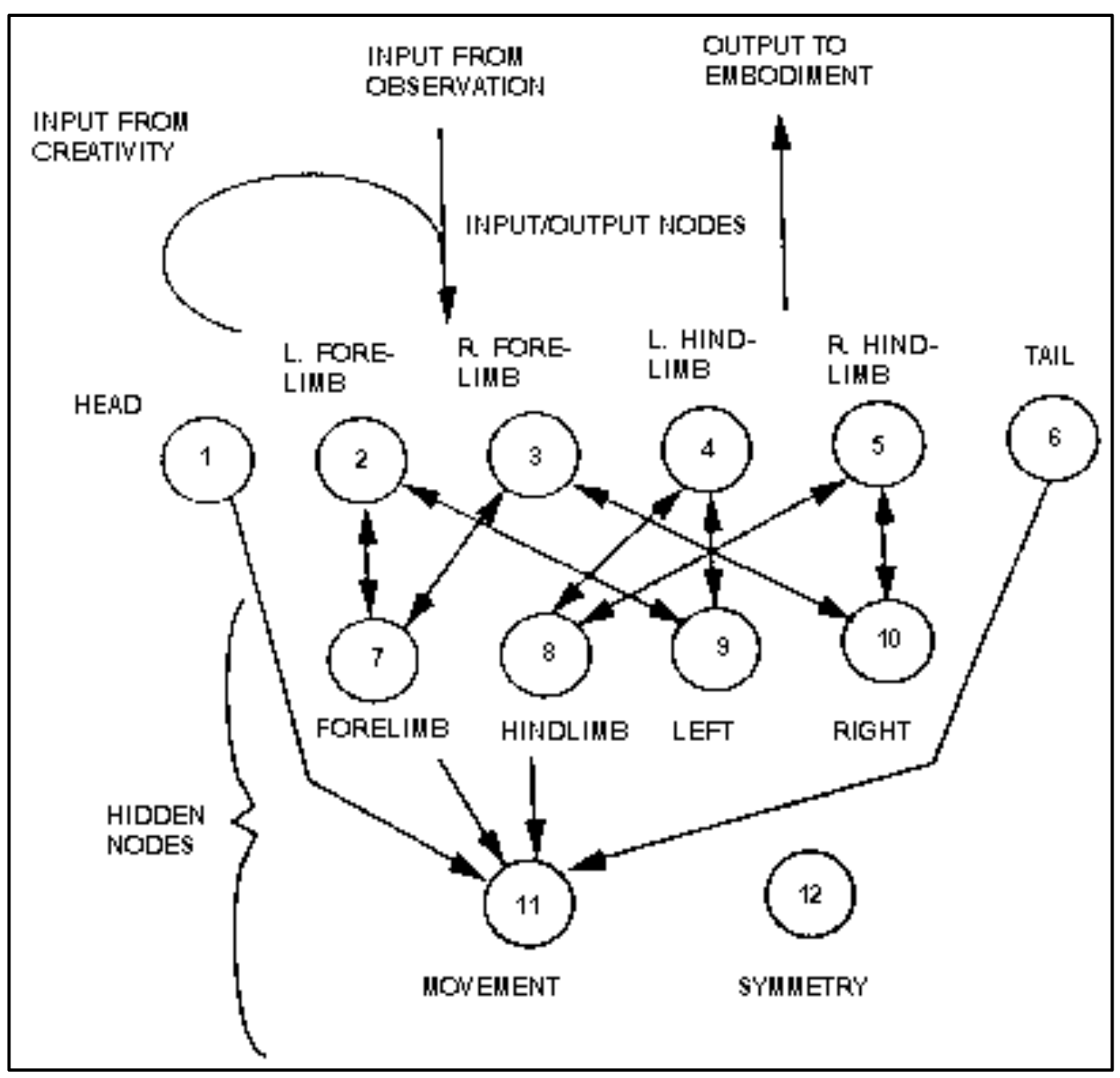

Figure 1. The neural network. Arrows represent connections with positive weights. For clarity, negative connections and connections to the symmetry unit are not shown.

Hidden units are linked with positive weights to input/output units that are positive instances of the concepts they represent, and linked with negative weights to input/output units that represent negative instances of the ideas they represent (thus "left arm" is positively linked to "left" and negatively linked to "right"). Hidden units that represent opposite concepts have negative connections between them. The hidden units enable the network to encode the semantic structure of an idea, and their activations are used to bias the generation of variation.

The neural network starts with small random weights between input/output nodes. Weights between hidden nodes, and weights between hidden nodes and input/output nodes, are fixed at +/- 1.0. Patterns (representing ideas) are learned by training for 50 iterations using the generalized delta rule (Rumelhart et. al. 1986) with a sigmoid activation function (Hinton 1981). The basic idea is that patterns are repeatedly presented to the network, and each time, the network's actual output is compared to the desired output. Since this neural network is an autoassociator, we desire simply that the output be identical to the input. The relevant variables are:

$a_j$ = activation of unit $j$
$t_j$ = the $j$th component of the target pattern (the external input)
$\omega_{ij}$ = weight on line from unit $i$ to unit $j$
$\beta = 0.15$
$\theta = 0.5$

$$a_j = \frac{1}{\left(1 + e^{-\beta\left[\sum w_{ij} a_i + \theta\right]}\right)} \qquad (1)$$

For the movement node, we use the absolute value of $a_i$ (since negative movement is not possible; the least you can move is to not move at all).

The comparison between input and output involves computing an error term, which is used to modify the pattern of connectivity in the network such that its responses become more correct. The error signal, $\delta_j$, is calculated such that the more intermediate (and thus 'uncommitted's) the activation of the unit, the more it is affected. For input/output units it is computed as follows:

$$\delta_j = (t_j - a_j) a_j (1 - a_j) \qquad (2)$$

For hidden units we do the same thing by determining how much each unit is contributing to the error as follows:

$$\delta_i = a_j (1 - a_j) \sum \delta_j w_{ij} \qquad (3)$$

**1.3 The Embodiment**

The embodiment is a six-digit array specifying the behavior of the six body parts. While the output of the neural network represents what the agent is thinking about, the embodiment represents what it is actually doing. An idea cannot be observed and imitated by other agents until it has been copied from the neural network to the embodiment and is implemented as an action.

**1.4 The Fitness Function**

An optimal action is one where all body parts except the head are moving, and limb movement is anti-symmetrical. (Thus if the left arm is moving up, the right arm is moving down, and vice versa.) This is implemented as follows:

F = fitness
$\mu = 2.5$
$a_m$ = activation of movement hidden node
$a_s$ = activation of symmetry hidden node
$a_h$ = activation of head node

$i = 1$ if $a_h = 0.0$, otherwise $i = 0$

$$F = \mu a_m + 2\mu a_s + \mu i \qquad (4)$$

This fitness function corresponds to a relatively realistic display, but it also has some interesting properties. An agent that develops the general rule "movement improves fitness" risks overgeneralization since head stability contributes as much to fitness as movement at every other limb. This creates a situation that is the cultural analog of overdominance in genetics; the optimal value of this locus lies midway between the two extremes. We also have a situation analogous to bidirectional selection or underdominance; the optimal value of the tail locus lies at either of the two extremes. (The desire to investigate underdominance was the reason for giving the agents tails). There is a cultural analog of epistasis-where the fitness at one locus depends on which allele is present at another locus. Epistasis is present because the value of what one limb is doing depends on what its counterpart is doing; for example, if the left leg is moving backward the right should be moving forward, and vice versa. Finally, since there is one optimal allele for the head, two optimal alleles for the tail, two optimal arm combinations, and two optimal leg combinations, we have a total of eight different optimal actions. This enables us to perform a comparative analysis of diversity under different ratios of creation to imitation.

**1.5 Using Experience to Bias the Generation of Novelty**

The idea here is to translate knowledge acquired during evaluation of an action into educated guesses about what increases fitness. Each locus starts out with the allele for no movement, and with an equal probability of mutating to each of the other two alleles (those for upward and downward movement). A new action is not learned unless it is fitter than the currently-implemented action, so we use the difference between the two to bias the direction of mutation.

Two rules of thumb are used. The first rule is: if the fitter action codes for more movement, increase the probability of forward mutation and decrease the probability of back mutation. Do the opposite if the fitter action codes for less movement. This rule of thumb is based on the assumption that movement in general (regardless of which particular body part is moving) can be beneficial or detrimental. This seems like a useful generalization since movement of any body part uses energy and increases the likelihood of being detected. It is implemented as follows:

$a_{m1}$ = activation of movement unit for currently-implemented meme
$a_{m2}$ = activation of movement unit for new meme
$p(fmut)_i$ = probability of forward mutation at allele $i$ (increased movement)
$p(bmut)_i$ = probability of backward mutation at allele $i$ (decreased movement)

IF ($a_{m2} > a_{m1}$)
THEN $p(fmut)_i = \text{MAX}(1.0, p(fmut)i + 0.1)$
ELSE IF ($a_{m2} < a_{m1}$)
THEN $p(fmut)_i = \text{MIN}(0.0, p(fmut)_i - 0.1)$

$p(bmut)_i = 1 - p(fmut)_i$

The second rule of thumb biases the agent either toward or away from symmetrical limb movement. It has two parts. First, if in the fitter action both members of one pair of limbs are moving either up or down, increase the probability that you will do the same with the other pair of limbs. Second, if in the fitter action, one member of a pair of limbs is moving in one direction and its counterpart is moving in the opposite direction, increase the probability that you will do the same with the other pair of limbs. This generalization is also biologically useful, since many beneficial behaviors (walking, *etc*.) entail movement of limbs in opposite directions, while others (galloping, etc.) entail movement of limbs in the same direction. The implementation of this rule is analogous to that of the first rule.

In summary, each action is associated with a measure of its effectiveness, and generalizations about what seems to work and what does not are translated into guidelines that specify the behavior of the cultural algorithm.

## 2. PROTOCOL

Agents are in a two-dimensional wrap-around 10x10 grid-cell world, one agent per cell. Each iteration, every agent has the opportunity to (1) acquire an idea for a new action, either by imitating a neighbor, or creating one anew (2) update the mutation operator, and (3) implement the new action.

Agents have an equal probability of creating and imitating. To invent or create a new idea, the cultural algorithm is applied to the idea currently represented on the input/output layer of the neural network. For each locus, the agent decides whether mutation will take place. The probability of mutation is specified globally at the beginning of a run. If it decides to mutate, the direction of mutation is stochastically determined. If the new idea has a higher fitness than the currently-implemented idea, the agent learns and implements the action specified by that idea.

To acquire an idea through imitation, an agent randomly chooses one of its eight neighbors and evaluates the fitness of the action the neighbor is implementing. If its own action is fitter than that of the neighbor, it chooses another neighbor, until it has either observed all eight neighbors, or found one with a fitter action. If no fitter action is found, the agent does nothing. Otherwise, the neighbor's action is copied to the input/output layer, and it is learned and implemented.

Since in both creation and imitation, a new idea is not acquired unless it is fitter than the currently implemented action, the new idea provides information that is used by the cultural algorithm. For example, since we arbitrarily chose a fitness function in which movement is generally beneficial, if the new action does code for more movement than the old one, the probability of forward mutation will almost always increase.

No matter how the new idea has been acquired, it gets implemented as an action when it is copied from the neural network to the embodiment. In the 'no mental simulation' condition, whether the new idea was acquired through creation or imitation, it must be implemented as an action for at least one iteration before its fitness can be assessed. In this case, mutation operators are updated the following iteration.

## 3. RESULTS

The following experiments were conducted using a mutation rate of 0.17 per locus, a 1:1 creation to imitation ratio, and all cultural evolution strategies operative, unless otherwise indicated.

### 3.1 Outline of a Run: Culture Evolves

Initially all agents were immobile, thus the number of different actions implemented was zero, as shown in Figure 2. The immobility idea quickly mutated to a new idea that coded for movement of a body part. This new idea had a higher fitness and was preferentially implemented. As ideas continued to be created, get implemented as actions, and spread through imitation, the society evolved toward increasingly fit actions.

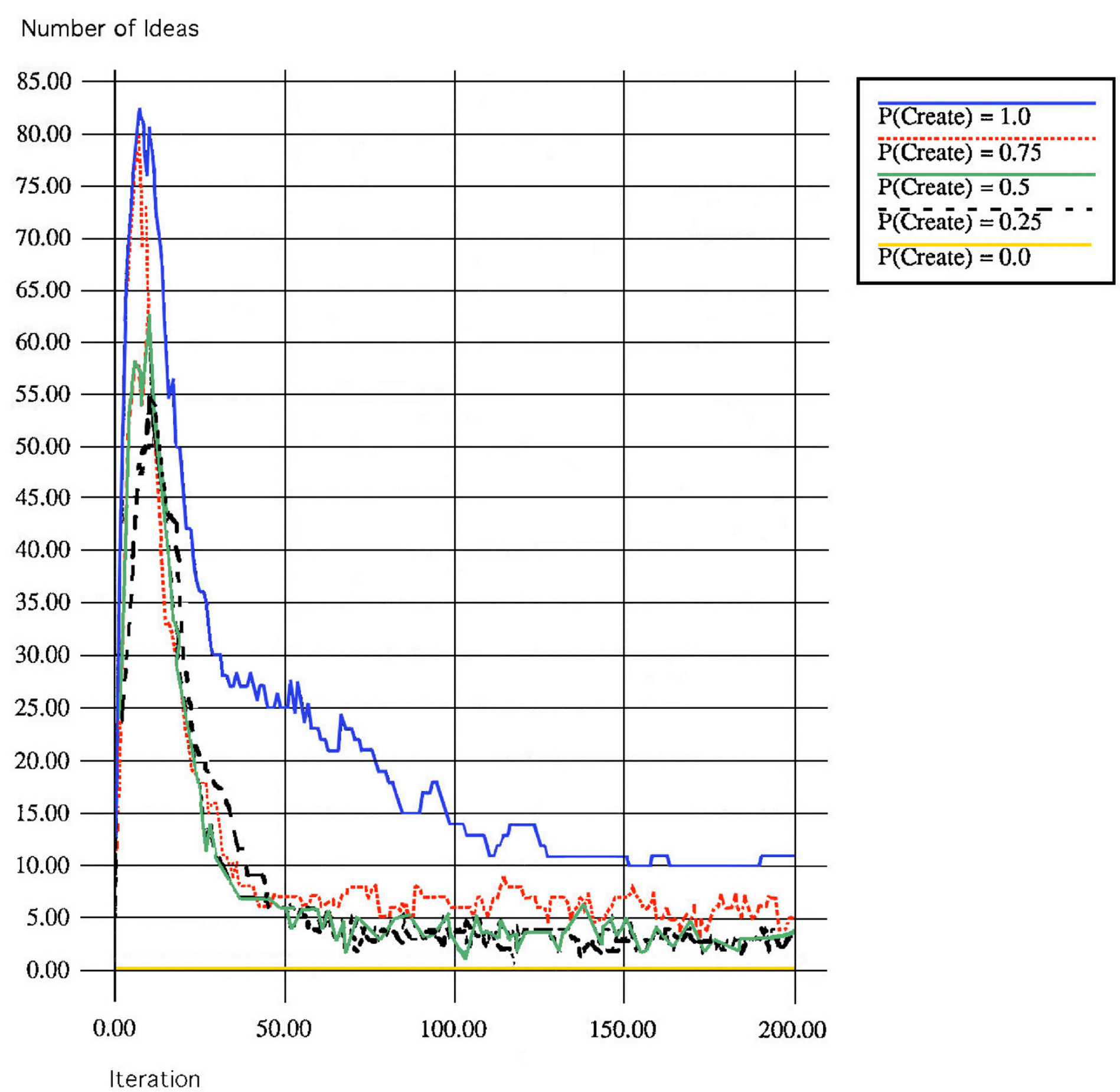

Figure 2. Effect of varying p(create) to p(imitate) ratio on number of different ideas.

### 3.2 Trade-off Between Diversity and Global Optimization

Figure 2 shows how diversity peaked when the first maximally fit idea was found, and decreased as the society converged on maximally-fit ideas. As in a genetic algorithm, increasing the frequency of variation-inducing operations-in this case, the creativity to imitation ratio-increased diversity. This was true both as the society was evolving, and when it finally stabilized. An interesting result can be seen if one looks more closely at how diversity varied with the creation to imitation ratio. Diversity ranged from 1-2 actions when p(create) = 0.25, to 10-11 actions when p(create) = 1.0. When p(create) = 0.75, the society converged on 7-8 actions. Thus it found all (or nearly all) of the fittest actions. A nice balance was thereby struck between the diversifying effect of creation and the converging effect of imitation.

### 3.3 Frequency of Change Must be Intermediate

Agents could vary with respect to not only just the frequency with which they invented, but with respect to how much change they introduced when they did. As in a genetic algorithm, evolution did not occur in the complete absence of mutation. The best performance was obtained with between 0.07 and 0.22 mutations per locus (Figure 3). In biological terms this would constitute a very high mutation rate, however because there are so few loci it ends up being approximately one mutation per innovation event, which is very reasonable. There are however two reasons why it can afford to be high. The first is that mental simulation ensures that unfit ideas are not implemented as actions. The second is that fit actions are imitated by others, thus never lost from the society as a whole.

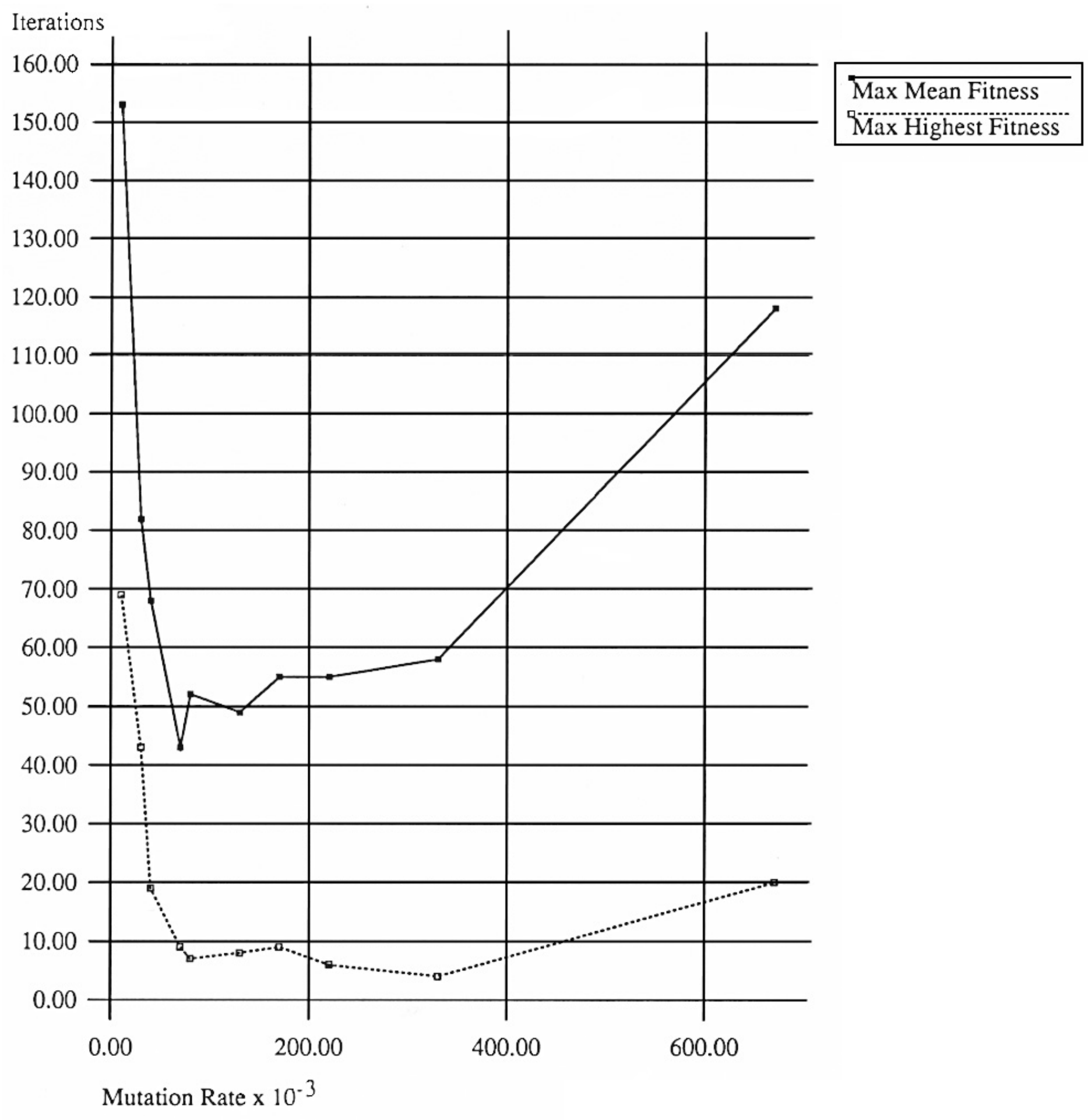

Figure 3. Optimization time decreases sharply, and then increases slowly, as mutation rate increases. This trend holds true for both the mean fitness of all ideas and the fittest idea that has appeared in a given iteration.

Note that this is similar to what happens in a genetic algorithm when the mutation rate is much above the minimum necessary for evolution. In both MAV and the GA, the frequency of change must be intermediate between the two extremes.

### 3.4 Epistasis Decreases Rate of Fitness Increase

As in biology, fitness increased more slowly, and stabilization took longer, for epistatically linked loci than for either over- or underdominant loci Figures 4 and 5 show this at two different mutation rates; in fact, it was observed at every mutation rate tested. In figure 4, we see that whereas the over-dominant locus had stabilized by the 100th iteraction, and the under-dominant by the 150th, the two epistatic loci took 200 iterations to stabilize. In figure 5 whereas the overdominant locus stabilizes by the 70th iteration, and the under-dominant by approximately iteration 130, the two epistatic loci never do manage to stabilize. The phenomenon can be attributed to the fact that for epistatically linked loci there are more constraints to be met; what one arm should be doing, for example, depends on what the other arm is doing. It is also interesting to note that the epistatic loci to some extent mirror one another, such that if the mean activation for one locus increases, the mean activation for the epistatically linked locus decreases, and vice versa.

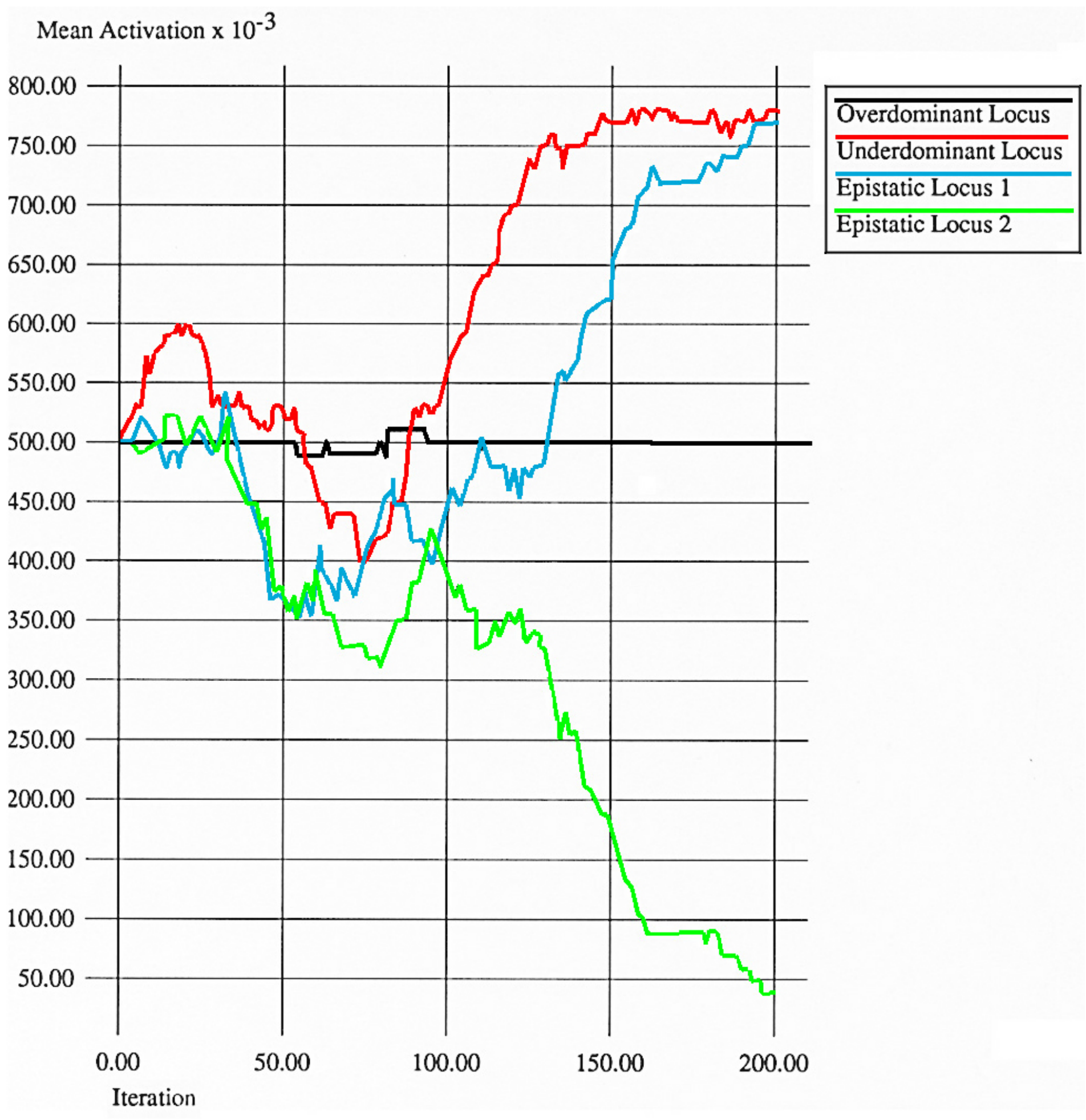

Figure 4. Mean loci activations with a mutation rate of 0.01. Over-dominant locus stabilizes most quickly, followed by under-dominant locus, and then epistatic loci.

### 3.5 Cultural Drift

Since we have eight optimal ideas, there are many stable configurations for the distribution of ideas. Figures 4 and 5 reveal amongst equally-fit alleles the presence of drift -- the term biologists use to refer to changes in the relative frequencies of different alleles due to random sampling processes in a finite population (Wright, 1969). Drift is indicated by the fact that since we are looking at mean activation values across the entire society, if the activation value is very high or very low that means that almost all agents had stabilized on the same value at a particular locus. This is the case for both the over-dominant and epistatic loci. (For the under-dominant locus it is not possible to distinguish between all individuals stabilized on the intermediate value, or as many stabilized above this value as below.) This is in accord with Cavalli-Sforza and Feldman's (1981) mathematical model of culture.

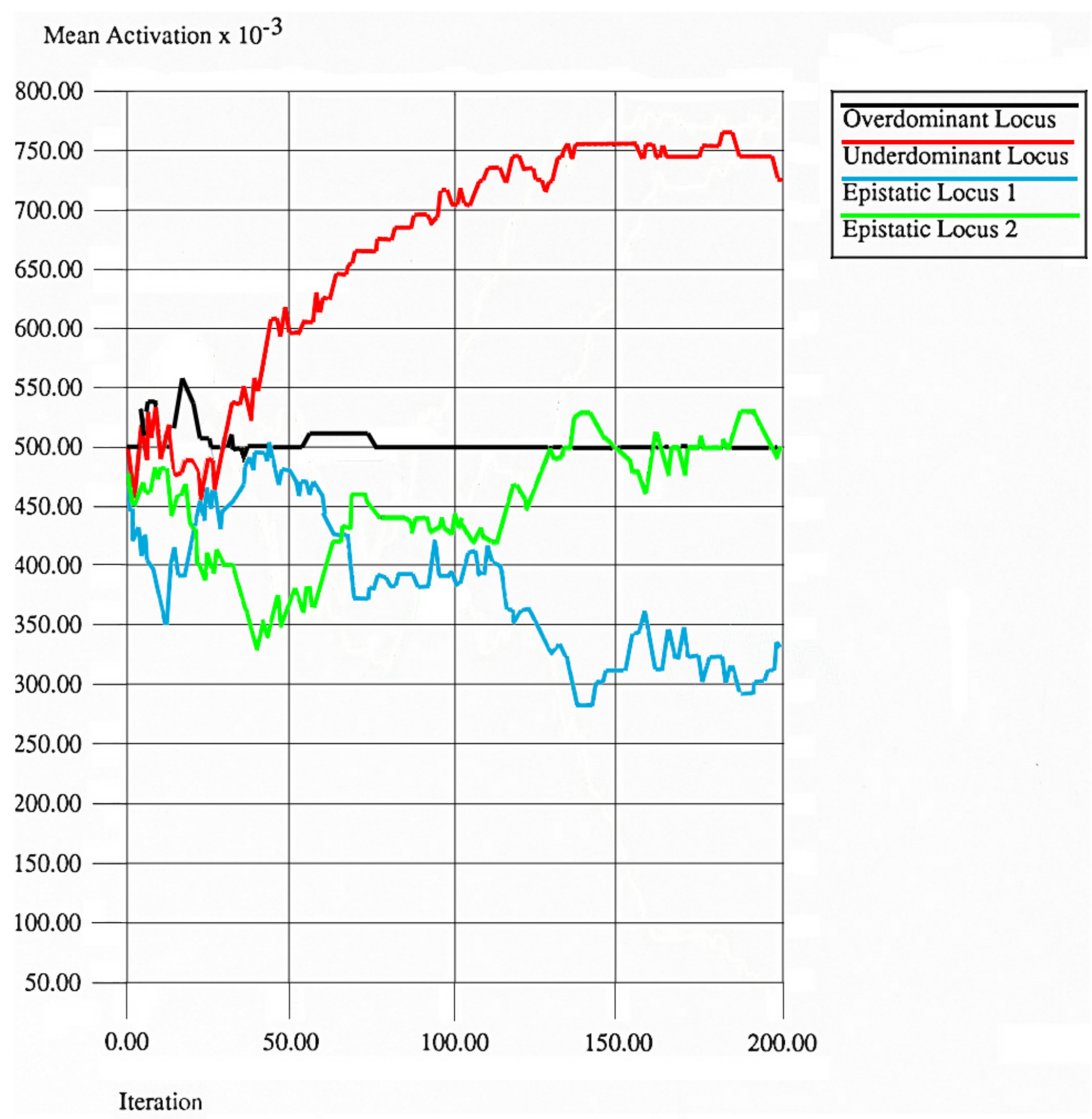

Figure 5. Mean loci activations with a mutation rate of 0.67. Over-dominant locus stabilizes most quickly. Under-dominant locus takes longer. After 200 iterations, epistatic loci still have not stabilized.

### 3.6 Effect of Knowledge-based Operators, Imitation, and Mental Simulation

The three cultural evolution strategies-knowledge-based operators, imitation, and mental simulation-were made inoperative one at a time to determine their contribution to optimization speed and peak mean fitness. These experiments were performed separately, and also cumulatively, adding each strategy one at a time. Since, the results were comparable for the separate and cumulative experiments, only the cumulative results are presented here (Figure 6).

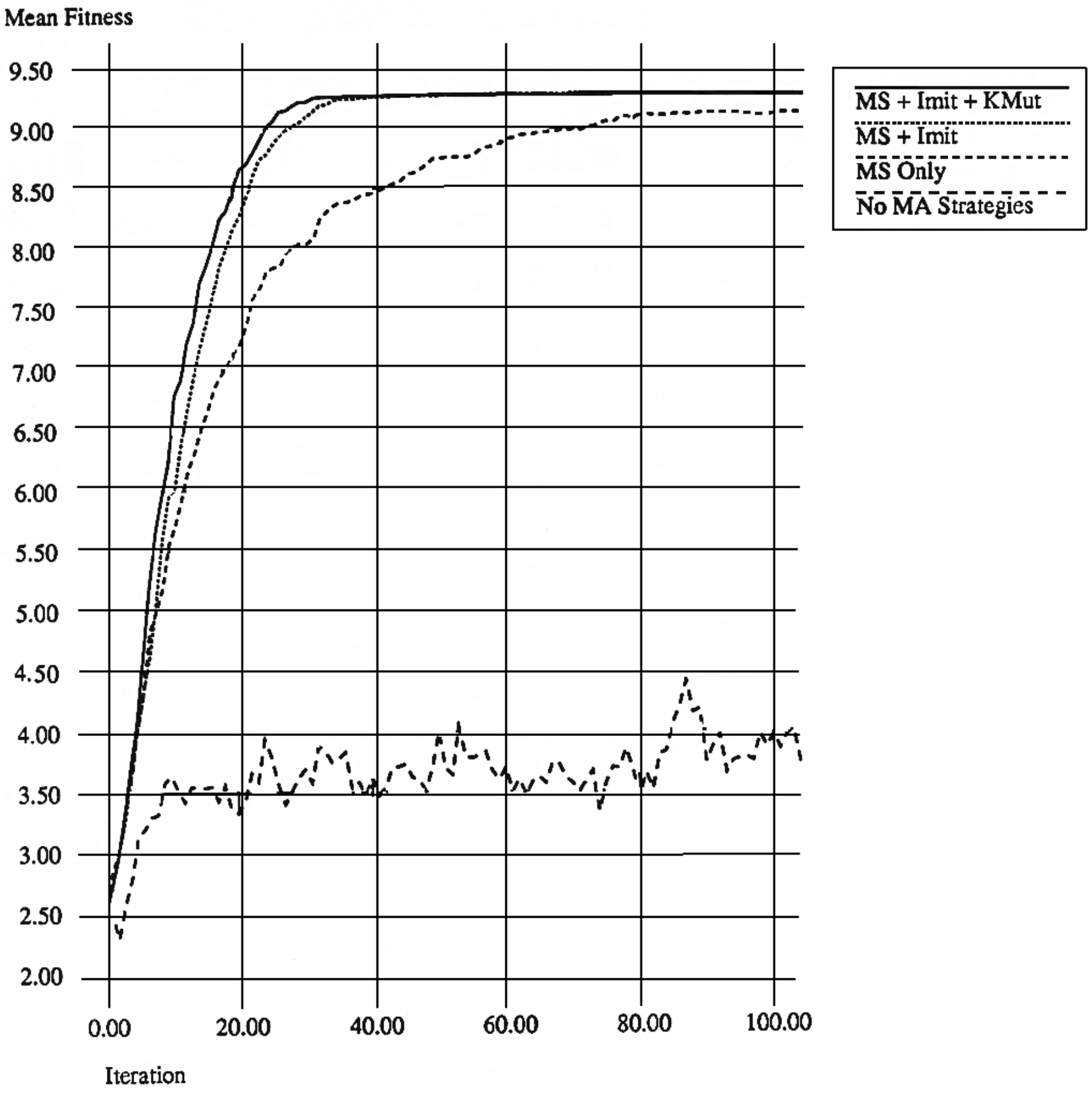

Figure 6. Cumulative improvement with successively applied cultural strategies. MS refers to mental simulation. 'Imit' refers to imitation. 'Kmut' refers to knowledge based operators.

All three cultural evolution strategies increased the rate of optimization. Mental simulation and imitation also increased peak mean fitness.

### 3.7 Fittest Society with Creation to Imitation Ratio of 2:1

The highest mean fitness was achieved when both creation and imitation were employed, as illustrated in Figure 7. The best performance was observed when the creativity to imitation ratio was either 1:1 or 3:1. The society with a 3:1 ratio improved most quickly until the 21st iteration, at a mean fitness of 8.6. The society with a 1:1 ratio then overtook it and converged entirely on optimal actions slightly earlier than the 3:1 society (32 iterations as opposed to 37). Thus it can be said that overall the optimal creativity to imitation ratio is most likely midway between these values, or approximately 2:1. The 1:3 ratio society took longer to converge, though it did eventually after 47 iterations. The society that just created and never imitated had the opposite problem. It did as well as the 1:1 and 3:1 ratio societies for the first ten iterations or so, at which a mean fitness of 6.2 was attained. However, after that, its performance dropped compared to the others, such that even after 100 generations it never did converge on optimal solutions. As might be expected, when there was no creativity, just imitation, then there was no evolution at all; the agents simply remained immobile.

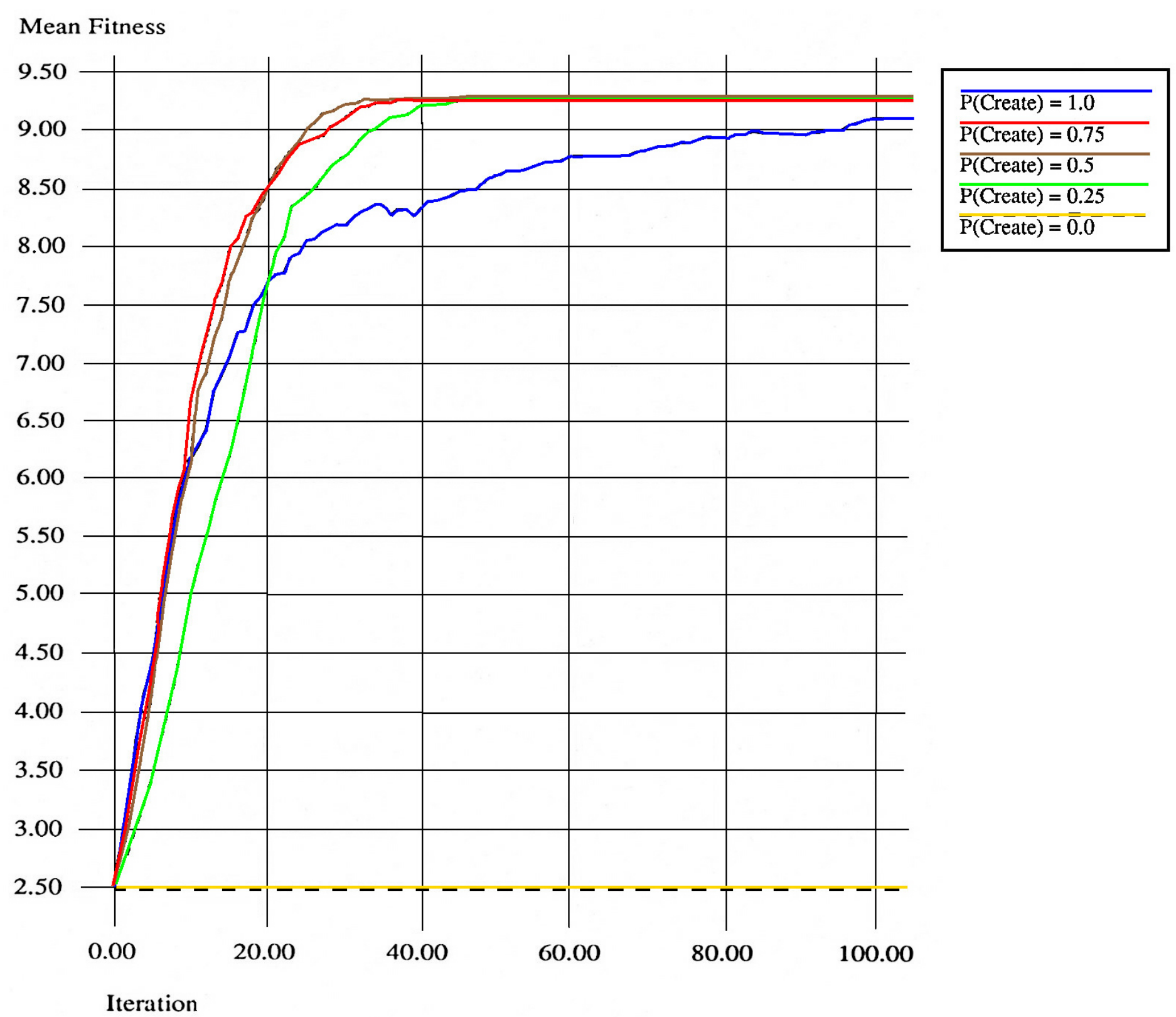

Figure 7. Effect of varying the creation-to-imitation ratio. The optimum seems to be between 0.5 and 0.75.

However, it is interesting to note that the fitness of the fittest idea (Figure 8) increased as a function of the ratio of creation to imitation. Since the agents with the fittest ideas gain nothing by imitating others, there is a trade-off between average action fitness, and fitness of the fittest action. (Of course, this result should not be taken too seriously as indicating that smart people don't need to imitate, since the agents in MAV only had one problem to solve. Thus, those who happened to be lucky simply got a head start.)

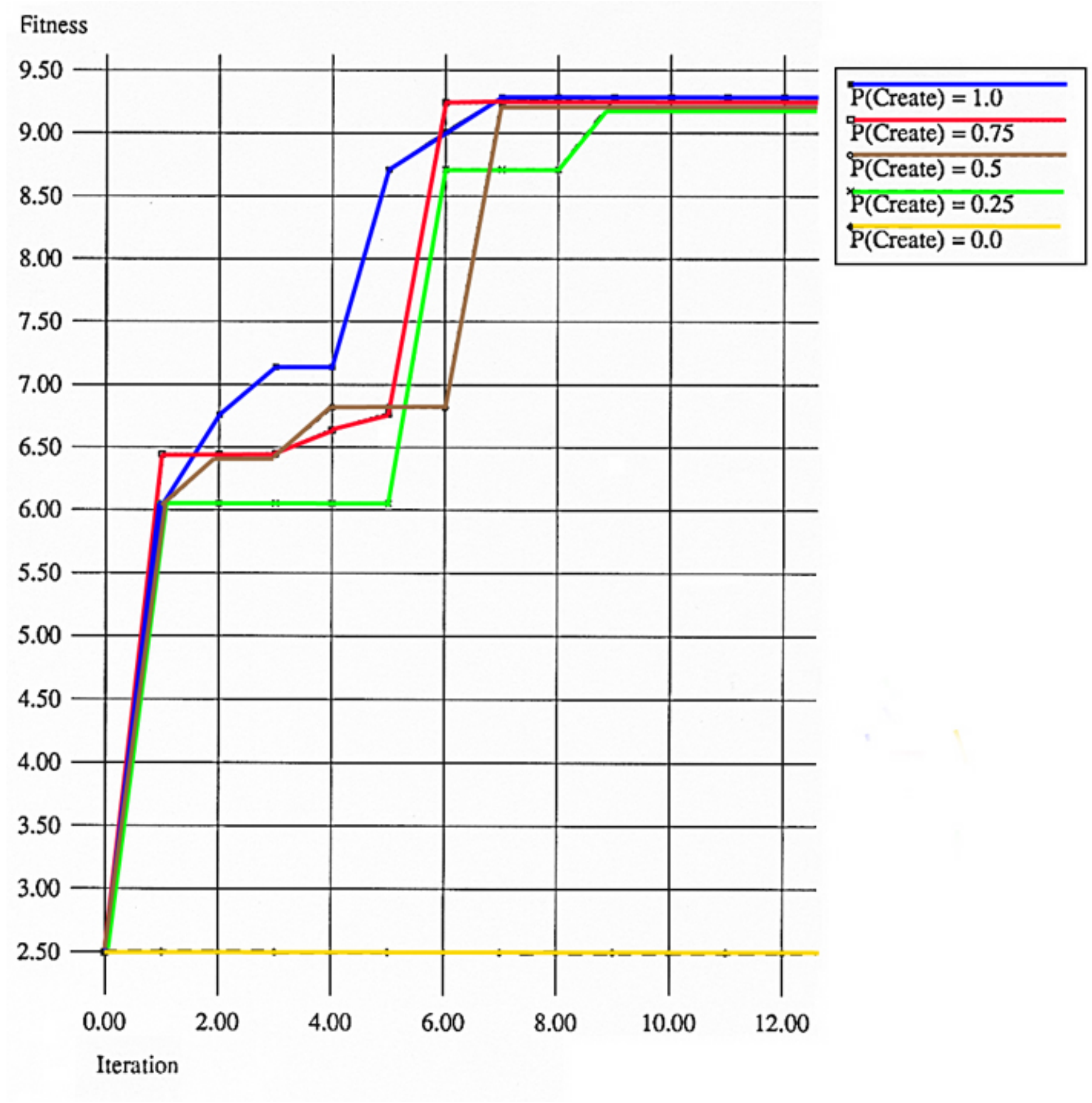

Figure 8. Effect of varying p(create) to p(imitate) ratio on the fitness of the fittest idea.

## 4. COMPARISON WITH OTHER APPROACHES

The computational approach taken here allows us to look for patterns that arise over time when the cultural activities of inventing and imitating are carried out in parallel in a society of interacting individuals. To my knowledge, MAV is the first computational model of the evolution of culture as opposed to the effect of culture on biological evolution. For example, Hutchins and Hazelhurst (1992) used a computer model to explore the relationship between environment, internal representation of the environment, and cultural artifacts that mediate the transmission of knowledge about environmental regularity vertically across generations. In MAV, on the other hand, we stick to one generation so that the effects of cultural evolution can be disentangled from biological evolution. Similarly, computer models of the evolution of creativity (Sims, 1991; Todd & Latham, 1992) and cooperation (Axelrod, 1985), although they explore a cultural process, they use a genetic algorithm-a model of biological evolution. Axelrod's work has inspired others, who have taken a more culturally realistic approach (*e.g.*, Conte & Paolucci forthcoming; Hales forthcoming), but in these studies the space of possible cultural entities is too small to evolve, so the study is really a study of how particular cooperative strategies diffuse across a population of agents. Note, however, that although MAV is modeled after cultural evolution, even it is too simple to explore many cultural phenomena. The space of possible ideas is fixed and small, and (unlike real life) the fitness function is predetermined and static. It would be particularly interesting to explore the effects of a dynamically changing fitness function given recent findings that maintaining diversity in a population is, over the long term, more important than global fitness in the short term (Hutter, 2001). Another shortcoming of MAV, imitation and innovation are not as discrete in real life as MAV would suggest. Despite these shortcomings, however, MAV demonstrates the feasibility of computationally modeling the processes by which creative ideas spread through a society giving rise to observable patterns of cultural diversity.

## 5. SUMMARY

If culture, like biology, is a form of evolution, it should be possible to abstract the underlying skeleton of the process and develop a minimal model of it analogous to the genetic algorithm. MAV is a minimal computer model of the process by which culture evolves. It consists of an artificial society of neural network-based agents that don't have genomes, and neither die nor have offspring, but they invent, imitate, and implement ideas, and thereby their actions gradually become more fit. Every iteration, each neural-network based agent in an artificial society has the opportunity to acquire a new idea, either through 1) innovation, by mutating a previously learned idea, or 2) imitation, by copying an action performed by a neighbor.

The program exhibits features observed in biology such as: (1) drift, (2) epistasis increases time to reach equilibrium, (3) increasing the frequency of innovative or variation-generating operations increases diversity, and (4) although in the absence of variation-generating operations, culture does not evolve, increasing innovation much beyond the minimum necessary for evolution causes average fitness to decrease.

The model also addresses the evolutionary consequences of phenomena specific to culture. Imitation, mental simulation (ability to assess the relative fitness of an action before actually implementing it), and strategic (as opposed to random) generation of variation all increase the rate at which fitter actions evolve. The higher the ratio of innovation to imitation, the greater the diversity, and the higher the fitness of the fittest action. Interestingly, however, for the society as a whole, the optimal innovation-to-imitation ratio was approximately 2:1 (but diversity is then compromised). For the agent with the fittest behavior, the less it imitated (i.e. the more computational effort reserved for innovation), the better.